\documentclass[showpacs]{revtex4} 
\usepackage{epsfig,amsmath,amssymb,graphicx,upgreek,textcomp}
\usepackage{natbib}
\usepackage[english]{babel} %

\begin{document}

\vspace{0mm}
\title{Quantum relativistic equation for a probability amplitude} %
\author{Yu.M. Poluektov}
\email{yuripoluektov@kipt.kharkov.ua (y.poluekt52@gmail.com)} %
\affiliation{National Science Center ``Kharkov Institute of Physics and Technology'', 61108 Kharkov, Ukraine} %

\begin{abstract}
The relativistic quantum equation is proposed for the complex wave
function, which has the meaning of a probability amplitude. The
Lagrangian formulation of the proposed theory is developed. The
problem of spreading of a wave packet in an unlimited space is
solved. The relativistic correction to the energy levels of a
harmonic oscillator is found, leading to a violation of their
equidistance.
\newline%
\end{abstract}
\maketitle

\section{Introduction}\vspace{-0mm} 
Fundamental issues related to the completeness of the description of
reality in quantum mechanics were discussed long ago in well-known
works [1,2] and then in many others. These discussions continue in
present time. The Schr\"{o}dinger equation for the complex wave
function $\psi$, which, according to Born's interpretation, has the
meaning of a probability amplitude, is nonrelativistic. This means
that the theory allows propagation of signals at arbitrarily high
speeds. Perhaps, some of the discussed difficulties and paradoxes of
quantum mechanics are related to this circumstance. It should also
be noted that quantum theory is used in the description and
interpretation of experiments with photons, which are relativistic
objects. So far there is no generally accepted relativistic equation
for the field, which could be interpreted as a probability amplitude
and would allow a Born probabilistic interpretation. The
probabilistic interpretation of the complex field within the
framework of Dirac theory was discussed in works [3,4].

In this paper the relativistic equation is proposed for the complex
scalar field, allowing its physical interpretation as a probability
amplitude. The theory is formulated within the framework of the
Lagrangian formalism. The spreading of a wave packet in an unlimited
space is considered. The relativistic correction has been found in
the theory of a harmonic oscillator, leading to a violation of the
equidistance of levels. The issue of the completeness of the quantum
description is not addressed in the paper.

\section{Relativistic equation for the complex scalar field}\vspace{-0mm} %
The recipe for the transition from the classical to the quantum
mechanical equation consists, as is known [5], in replacing momentum
and energy in the classical Hamiltonian $H={\bf p}^2/2m$ with the
operators ${\bf p}\rightarrow -i\hbar\nabla$ and $H\rightarrow
i\hbar\,\partial/\partial t$, which act on the complex function
$\psi$. In the relativistic case, the relationship between energy
and momentum is given by formula $H=\sqrt{m^2c^4+c^2{\bf p}^2}$,
where $m$ is the particle mass, $c$ is the speed of light. Using the
indicated substitution, we arrive at the relativistic equation
\begin{equation} \label{01}
\begin{array}{l}
\displaystyle{%
   i\hbar\frac{\partial\psi}{\partial t}=\sigma\sqrt{m^2c^4-\hbar^2c^2\Delta}\,\,\psi,  %
}
\end{array}
\end{equation}
where $\Delta$ is the Laplace operator, $\sigma=\pm 1$. This
equation is inconvenient for a number of reasons. Firstly, the time
and space coordinates enter here unequally, so that it does not have
an explicitly covariant form. Secondly, because in Eq.\,(1) the
Laplace operator appears under the sign of the root, and, therefore,
this equation is nonlocal. In order to get rid of the noted formal
shortcomings, the transition to the quantum mechanical equation is
usually carried out in the expression for the square of energy
$H^2=m^2c^4+c^2{\bf p}^2$, which leads to the Klein-Gordon-Fock equation %
\begin{equation} \label{02}
\begin{array}{l}
\displaystyle{%
  \bigg(\Delta-\frac{1}{c^2}\frac{\partial^2}{\partial t^2} - \lambdabar^{-2} \bigg)\psi=0,    %
}%
\end{array}
\end{equation}
where $\lambdabar\equiv\hbar\big/mc$\, is the Compton length. In the
following we will also use 4-dimensional notation, in which an
arbitrary 4-vector is $B\equiv B_\mu\equiv\big({\bf B}, B_4=iB_0\big)$, %
in particular $x\equiv x_\mu\equiv\big({\bf x}, x_4=ix_0=ict\big)$. %
The scalar product of two vectors is written as %
$AB\equiv A_\mu B_\mu = {\bf A}{\bf B}+A_4B_4 = {\bf A}{\bf B}-A_0B_0$. %
A summation from 1 to 4 is implied over repeating Greek indices, and
a summation from 1 to 3 -- over repeating Latin indices. In this
notation equation (2) can be written in the explicitly covariant form %
\begin{equation} \label{03}
\begin{array}{l}
\displaystyle{%
  \bigg(\frac{\partial^2}{\partial x_\mu^2} - \lambdabar^{-2} \bigg)\psi(x)=0.    %
}%
\end{array}
\end{equation}
From this equation there follows the continuity equation
\begin{equation} \label{04}
\begin{array}{l}
\displaystyle{%
  \frac{\partial \rho}{\partial t} +\nabla\cdot{\bf j} = 0,  %
}%
\end{array}
\end{equation}
where
\begin{equation} \label{05}
\begin{array}{l}
\displaystyle{%
  \rho=i\frac{\lambdabar}{2c}\bigg(\psi^*\frac{\partial\psi}{\partial t} - \psi\frac{\partial\psi^*}{\partial t} \bigg), \qquad    %
  {\bf j}=-i\frac{\hbar}{2m}\big(\psi^*\nabla\psi - \psi\nabla\psi^* \big).     %
}%
\end{array}
\end{equation}
Here the quantity $\rho$ is not positively definite, and therefore
cannot be interpreted as a probability density. In addition, the
transition to a higher order equation leads to the emergence of new
solutions. In connection with this, the problem arises of
constructing the quantum relativistic equation for the complex
field, which would have the meaning of a probability amplitude.

\section{Relativistic quantum equation for a probability amplitude}\vspace{-0mm} %
In spite of the shortcomings noted above, let us consider equation
(1) as the equation for the probability amplitude $\psi(x)$. The
advantage of this equation, which to a great extent compensates for
the difficulties noted above, consists in the possibility of
interpreting the complex field as a probability amplitude. In this
equation the sign in front of the root can be chosen arbitrarily,
since this will not affect the physical results, but only affect the
form of the time dependence of the wave function for stationary
states $\psi\sim\exp\!\big(\!-i\sigma Et/\hbar\big)$. In order for
equation (1) to have the generally accepted form of the
Schr\"{o}dinger equation in the nonrelativistic limit, we will
assume in (1) $\sigma=+1$. The equation
\begin{equation} \label{06}
\begin{array}{l}
\displaystyle{%
   i\hbar\frac{\partial\psi}{\partial t}=\sqrt{m^2c^4-\hbar^2c^2\Delta}\,\,\psi  %
}
\end{array}
\end{equation}
does not have an explicitly covariant form, but leads to the
continuity equation for the probability density $|\psi|^2$: %
\begin{equation} \label{07}
\begin{array}{l}
\displaystyle{%
  \frac{\partial |\psi|^2}{\partial t} +\nabla\cdot{\bf j} = 0,  %
}%
\end{array}
\end{equation}
where
\begin{equation} \label{08}
\begin{array}{l}
\displaystyle{%
  \nabla\cdot{\bf j}=i\frac{mc^2}{\hbar}\Big(\psi^*\sqrt{1-\lambdabar^2\Delta}\,\,\psi-\psi\sqrt{1-\lambdabar^2\Delta}\,\,\psi^*\Big), %
}%
\end{array}
\end{equation}
and ${\bf j}$ has the meaning of the probability flux density. The
continuity equation (7) for the probability density can be written
in 4-dimensional form
\begin{equation} \label{09}
\begin{array}{l}
\displaystyle{%
  \frac{\partial j_\mu(x)}{\partial x_\mu} = 0,  %
}%
\end{array}
\end{equation}
where $j_\mu\equiv\big({\bf j}, ic|\psi|^2\big)$ is a 4-vector of
flux density, so that it has the same form in any inertial reference
system. Thus, the equation for the probability amplitude (6),
written in an arbitrary reference system, leads to the Lorentz
covariant form of the probability conservation law (9). Since under
the Lorentz transformation $x_\mu\rightarrow x_\mu'=a_{\mu\nu}x_\nu$, %
where $a_{\mu\nu}a_{\mu\nu'}=\delta_{\nu\nu'}$, the 4-vector is
transformed according to the law $j_\mu'(x')=a_{\mu\nu}j_\nu(x)$, %
then the probability density and the probability flux density in
different systems are connected by the relations
\begin{equation} \label{10}
\begin{array}{ccc}
\displaystyle{%
  ic|\psi'(x')|^2 =  ic|\psi(x)|^2 + a_{4m}j_m(x), %
}\vspace{3mm}\\ %
\displaystyle{\hspace{0mm}%
  j_k'(x') = a_{k4}\,ic|\psi(x)|^2 + a_{km}j_m(x). %
}%
\end{array}
\end{equation}

Let us represent the complex field as the sum of the real and
imaginary parts
\begin{equation} \label{11}
\begin{array}{l}
\displaystyle{%
  \psi(x)=\frac{1}{\sqrt{2}}\big[\psi'(x) + i\psi''(x)\big].  %
}%
\end{array}
\end{equation}
Then equation (6) turns out to be equivalent to a system of two
equations for the real fields $\psi'$ and $\psi''$:
\begin{equation} \label{12}
\begin{array}{l}
\displaystyle{%
   \hbar\dot{\psi'}=\sqrt{m^2c^4-\hbar^2c^2\Delta}\,\,\psi'', \qquad \hbar\dot{\psi''}=-\sqrt{m^2c^4-\hbar^2c^2\Delta}\,\,\psi'.  %
}
\end{array}
\end{equation}
Here and henceforth we also use the notation $\dot{\psi}\equiv\partial\psi/\partial t$. %
One of the functions, for example $\psi''$, can be eliminated, and then we arrive %
at the explicitly covariant Klein-Gordon-Fock equation for the real part %
\begin{equation} \label{13}
\begin{array}{l}
\displaystyle{%
   \ddot{\psi'}-c^2\Delta\psi'+c^2\lambdabar^{-2}\psi' = 0. %
}
\end{array}
\end{equation}
Along with this, the evolution of the imaginary part is determined
by the second equation (12). Similarly, the Klein-Gordon-Fock
equation can be obtained for the imaginary part of the function (11). %

Let us consider the meaning of the expression $\sqrt{1-\lambdabar^2\Delta}\,\,\psi$, %
containing the square root of the Laplace operator. For $x^2<1$ the
following expansions are valid [6]
\begin{equation} \label{14}
\begin{array}{l}
\displaystyle{%
   \sqrt{1\pm x}=1-\sum_{n=1}^\infty(\mp)^na_nx^n, \quad a_n\equiv\frac{(2n-3)!!}{(2n)!!},  \quad  %
   a_1=\frac{1}{2}, \quad a_2=\frac{1}{8}, \quad a_3=\frac{3}{48}, %
}
\end{array}
\end{equation}
\vspace{-5mm}%
\begin{equation} \label{15}
\begin{array}{l}
\displaystyle{%
   (1\pm x)^{-1/2}=1+\sum_{n=1}^\infty(\mp)^nb_nx^n, \quad b_n\equiv\frac{(2n-1)!!}{(2n)!!},  \quad  %
   b_1=\frac{1}{2}, \quad b_2=\frac{1}{8}, \quad b_3=\frac{15}{48}, %
}
\end{array}
\end{equation}
where $(2n)!!=2\cdot4\cdot6\ldots2n=2^nn!$, $(2n+1)!!=1\cdot3\cdot5\ldots(2n+1)=\frac{2^{n+1}}{\sqrt{n}}\,\Gamma\big(n+3/2\big)$. %
With allowance for these representations, by the expression with the
square root from the operator we will understand the following expansion %
\begin{equation} \label{16}
\begin{array}{l}
\displaystyle{%
  \sqrt{1-\lambdabar^2\Delta}\,\,\psi=\bigg(1-\sum_{n=1}^\infty a_n\lambdabar^{2n}\Delta^{n}\bigg)\psi. %
}%
\end{array}
\end{equation}
This expansion is valid under the condition $|\lambdabar^2\Delta\psi|<1$, %
which is assumed to be satisfied in the following. This means that
one considers the states in which the wave function changes over
distances greater than the Compton length. The opposite
ultrarelativistic case can be considered in a similar way. It should
be noted, however, that it is of rather methodological interest,
since in this limit the possibility of  transformation of particles
should be taken into account, which can only be done by passing to
the representation of secondary quantization.

Taking (16) into account, we obtain a formula for the probability
flux density. The divergence (8) can be written in the form
\begin{equation} \label{17}
\begin{array}{l}
\displaystyle{%
  \nabla\cdot{\bf j}=-i\frac{mc^2}{\hbar}\sum_{n=1}^\infty a_n\lambdabar^{2n}\big(\psi^*\Delta^n\psi-\psi\Delta^n\psi^*\big). %
}%
\end{array}
\end{equation}
From here it follows that the probability flux density has the form
\begin{equation} \label{18}
\begin{array}{l}
\displaystyle{%
  {\bf j}=\sum_{n=1}^\infty{\bf j}^{(n)}, \quad  %
  {\bf j}^{(n)}=-ica_n\lambdabar^{2n-1}\sum_{\alpha=1}^n %
  \Big[\big(\Delta^{\alpha-1}\psi^*\big)\big(\nabla\Delta^{n-\alpha}\psi\big) - %
  \big(\Delta^{\alpha-1}\psi\big)\big(\nabla\Delta^{n-\alpha}\psi^*\big)\Big]. %
}%
\end{array}
\end{equation}
Everywhere it is assumed that the Laplacian to the zeroth power is
equal to unity $\Delta^0\equiv 1$.

\section{Lagrangian formalism}\vspace{-0mm} %
Equation (6) can be obtained within the Lagrange formalism, if we
choose the Lagrangian in the form
\begin{equation} \label{19}
\begin{array}{l}
\displaystyle{%
  L=\frac{\psi^*}{2}\Big(i\hbar\dot{\psi}-mc^2\sqrt{1-\lambdabar^2\Delta}\,\,\psi\Big)- %
    \frac{\psi}{2}\Big(i\hbar\dot{\psi}^*+mc^2\sqrt{1-\lambdabar^2\Delta}\,\,\psi^*\Big). %
}%
\end{array}
\end{equation}
Lagrangian (19) is a function of variables $\psi,\psi^*$\,, $\dot{\psi},\dot{\psi}^*$ %
and derivatives with respect to spatial coordinates
$\Delta\psi, \Delta^2\psi,\ldots\Delta^n\psi,\ldots\Delta\psi^*,\Delta^2\psi^*,\ldots\Delta^n\psi^*,\ldots\,.$ %
In this case, the Euler-Lagrange equations have the form
\begin{equation} \label{20}
\begin{array}{l}
\displaystyle{%
  \frac{\partial L}{\partial\psi}-\frac{\partial}{\partial t}\frac{\partial L}{\partial\dot{\psi}} + %
  \Delta\frac{\partial L}{\partial\Delta\psi}+\ldots+\Delta^n\frac{\partial L}{\partial\Delta^n\psi} +\ldots =0, %
}%
\end{array}
\end{equation}
and a similar equation with the replacement $\psi\rightarrow\psi^*$. %
A substitution of the Lagrangian (19) into (20) leads to Eq.\,(6). %

In the absence of time-dependent external fields, from the
requirement of invariance of the Lagrangian under the time shift
$t\rightarrow t+t_0$ and $\psi(t)\rightarrow\psi'(t+t_0)$, there
follows the continuity equation for the energy density
\begin{equation} \label{21}
\begin{array}{l}
\displaystyle{%
  \frac{\partial {\rm H}}{\partial t} + \nabla\cdot{\bf j}_E = 0, %
}%
\end{array}
\end{equation}
where the energy density
\begin{equation} \label{22}
\begin{array}{ccc}
\displaystyle{%
  {\rm H}\equiv \dot{\psi}\frac{\partial L}{\partial\dot{\psi}} +\dot{\psi}^*\frac{\partial L}{\partial\dot{\psi}^*}-L=  %
  \frac{mc^2}{2}\Big[\psi^*\sqrt{1-\lambdabar^2\Delta}\,\,\psi + \psi\sqrt{1-\lambdabar^2\Delta}\,\,\psi^*\Big]. %
}%
\end{array}
\end{equation}
The energy flux density is given by the formulas
\begin{equation} \label{23}
\begin{array}{ccc}
\displaystyle{%
  {\bf j}_E=\sum_{n=1}^\infty{\bf j}_E^{(n)}, %
}\vspace{1mm}\\ %
\displaystyle{\hspace{0mm}%
  {\bf j}_E^{(n)}=\sum_{\alpha=1}^n %
  \Big[\big(\Delta^{n-\alpha}\dot{\psi}\big)\big(\nabla\Delta^{\alpha-1}L_n\big) - %
  \big(\nabla\Delta^{n-\alpha}\dot{\psi}\big)\big(\Delta^{\alpha-1}L_n\big) + %
}\vspace{2mm}\\ %
\displaystyle{\hspace{15mm}%
  +\big(\Delta^{n-\alpha}\dot{\psi}^*\big)\big(\nabla\Delta^{\alpha-1}L_n^*\big) - %
  \big(\nabla\Delta^{n-\alpha}\dot{\psi}^*\big)\big(\Delta^{\alpha-1}L_n^*\big) \Big]. %
}%
\end{array}
\end{equation}
Here we used the notation
\begin{equation} \label{24}
\begin{array}{ccc}
\displaystyle{%
  L_n\equiv\frac{\partial L}{\partial\Delta^n\psi}=mc^2a_n\lambdabar^{2n}\frac{\psi^*}{2}, \quad  %
  L_n^*\equiv\frac{\partial L}{\partial\Delta^n\psi^*}=mc^2a_n\lambdabar^{2n}\frac{\psi}{2}, %
}\vspace{3mm}\\ %
\displaystyle{\hspace{0mm}%
  \Delta^\alpha L_n =mc^2a_n\lambdabar^{2n}\frac{\Delta^\alpha\psi^*}{2}, \quad %
  \Delta^\alpha L_n^* =mc^2a_n\lambdabar^{2n}\frac{\Delta^\alpha\psi}{2}. %
}%
\end{array}
\end{equation}
Taking (24) into account, we have the final expression for the
energy flux density
\begin{equation} \label{25}
\begin{array}{ccc}
\displaystyle{\hspace{0mm}%
  {\bf j}_E^{(n)}=\frac{1}{2}mc^2a_n\lambdabar^{2n}\sum_{\alpha=1}^n %
  \Big[\big(\Delta^{n-\alpha}\dot{\psi}\big)\big(\nabla\Delta^{\alpha-1}\psi^*\big) - %
  \big(\nabla\Delta^{n-\alpha}\dot{\psi}\big)\big(\Delta^{\alpha-1}\psi^*\big) + %
}\vspace{2mm}\\ %
\displaystyle{\hspace{32mm}%
  +\big(\Delta^{n-\alpha}\dot{\psi}^*\big)\big(\nabla\Delta^{\alpha-1}\psi\big) - %
  \big(\nabla\Delta^{n-\alpha}\dot{\psi}^*\big)\big(\Delta^{\alpha-1}\psi\big) \Big]. %
}%
\end{array}
\end{equation}

The condition of invariance of the Lagrangian with respect to
spatial translations ${\bf x}\rightarrow{\bf x}'={\bf x}+{\bf x}_0$ %
leads to the continuity equation for the momentum density:
\begin{equation} \label{26}
\begin{array}{ccc}
\displaystyle{\hspace{0mm}%
  \pi_i\,+\nabla_k\,\sigma_{ik} = 0,%
}%
\end{array}
\end{equation}
where the momentum density $\pi_i$ and the flux density of
$i$\,-\,component of momentum $\sigma_{ik}$ are given by the formulas %
\begin{equation} \label{27}
\begin{array}{ccc}
\displaystyle{\hspace{0mm}%
  \pi_i=-\nabla_i\psi^*\frac{\partial L}{\partial\dot{\psi}^*}-\nabla_i\psi\frac{\partial L}{\partial\dot{\psi}}=%
  -\frac{i\hbar}{2}\big(\psi^*\nabla_i\psi-\psi\nabla_i\psi^*\big), %
}%
\end{array}
\end{equation}
\vspace{-4mm} %
\begin{equation} \label{28}
\begin{array}{ccc}
\displaystyle{%
  \sigma_{ik}=\sum_{n=1}^\infty\sigma_{ik}^{(n)}, %
}\vspace{2mm}\\ %
\displaystyle{\hspace{0mm}%
  \sigma_{ik}^{(n)}=-\sum_{\alpha=1}^n %
  \Big[\big(\Delta^{n-\alpha}\nabla_i\psi\big)\big(\nabla_k\Delta^{\alpha-1}L_n\big) - %
       \big(\nabla_k\Delta^{n-\alpha}\nabla_i\psi\big)\big(\Delta^{\alpha-1}L_n\big)+ %
}\vspace{2mm}\\ %
\displaystyle{\hspace{22mm}%
  +\big(\Delta^{n-\alpha}\nabla_i\psi^*\big)\big(\nabla_k\Delta^{\alpha-1}L_n^*\big) - %
       \big(\nabla_k\Delta^{n-\alpha}\nabla_i\psi^*\big)\big(\Delta^{\alpha-1}L_n^*\big)\Big]. %
}%
\end{array}
\end{equation}
With allowance for (24), the last formula takes the form
\begin{equation} \label{29}
\begin{array}{ccc}
\displaystyle{\hspace{0mm}%
  \sigma_{ik}^{(n)}=-\frac{1}{2}mc^2a_n\lambdabar^{2n}\sum_{\alpha=1}^n %
  \Big[\big(\Delta^{n-\alpha}\nabla_i\psi\big)\big(\nabla_k\Delta^{\alpha-1}\psi^*\big) - %
       \big(\nabla_k\Delta^{n-\alpha}\nabla_i\psi\big)\big(\Delta^{\alpha-1}\psi^*\big)+ %
}\vspace{2mm}\\ %
\displaystyle{\hspace{37mm}%
   +\big(\Delta^{n-\alpha}\nabla_i\psi^*\big)\big(\nabla_k\Delta^{\alpha-1}\psi\big) - %
   \big(\nabla_k\Delta^{n-\alpha}\nabla_i\psi^*\big)\big(\Delta^{\alpha-1}\psi\big)\Big]. %
}%
\end{array}
\end{equation}

Lagrangian (19) is also invariant under the phase transformation
\begin{equation} \label{30}
\begin{array}{ccc}
\displaystyle{\hspace{0mm}%
  \psi(x)\rightarrow\tilde{\psi}(x)=\psi(x)e^{i\alpha}, %
}%
\end{array}
\end{equation}
where the parameter $\alpha$ does not depend on coordinates and
time. A consequence of this phase symmetry is the probability conservation law %
\begin{equation} \label{31}
\begin{array}{ccc}
\displaystyle{%
  \bigg(\psi\frac{\partial L}{\partial\dot{\psi}}-\psi^*\frac{\partial L}{\partial\dot{\psi}^*}\bigg)+ %
}\vspace{2mm}\\ %
\displaystyle{\hspace{0mm}%
  +\nabla_i\sum_{n=1}^\infty\sum_{\alpha=1}^n %
  \Big[\big(\nabla_i\Delta^{n-\alpha}\psi\big)\Delta^{\alpha-1}L_n \, - %
       \big(\Delta^{n-\alpha}\psi\big)\nabla_i\Delta^{\alpha-1}L_n \, - %
}\vspace{2mm}\\ %
\displaystyle{\hspace{25mm}%
  -\big(\nabla_i\Delta^{n-\alpha}\psi^*\big)\Delta^{\alpha-1}L_n^* \, + %
       \big(\Delta^{n-\alpha}\psi^*\big)\nabla_i\Delta^{\alpha-1}L_n^* \Big] = 0. %
}%
\end{array}
\end{equation}
This equation, with taking into account formulas (24), coincides
with the continuity equation (7) for the probability density which
is obtained directly from equation (6).

\section{Interaction with the electromagnetic field}\vspace{-0mm} %
The interaction with the electromagnetic field is enabled using the
well-known derivative substitution
\begin{equation} \label{32}
\begin{array}{ccc}
\displaystyle{\hspace{0mm}%
  \frac{\partial}{\partial x_\mu}\rightarrow\frac{\partial}{\partial x_\mu} +i\frac{e}{\hbar c}\,A_\mu, %
}%
\end{array}
\end{equation}
where $e=\pm|e|$, $A_\mu\equiv\big({\bf A}, A_4=i\Phi\big)$  is a
4-vector potential of the electromagnetic field. In
three-dimensional notation, the substitution (32) is equivalent to
the following substitutions:
\begin{equation} \label{33}
\begin{array}{ccc}
\displaystyle{\hspace{0mm}%
  \nabla\rightarrow\nabla+\frac{e}{\hbar c}\,{\bf A}, \qquad  %
  i\hbar\frac{\partial}{\partial t}\rightarrow i\hbar\frac{\partial}{\partial t}+e\Phi,%
}\vspace{2mm}\\ %
\displaystyle{\hspace{0mm}%
  \Delta\rightarrow\Delta_A\equiv\Delta+ \frac{e}{\hbar c}\,\nabla{\bf A} + %
  \frac{2e}{\hbar c}\,{\bf A}\nabla + \bigg(\frac{e}{\hbar c}\bigg)^{\!2}{\bf A}^2. %
}%
\end{array}
\end{equation}
As a result, we obtain the quantum relativistic equation for the
probability amplitude in the electromagnetic field:
\begin{equation} \label{34}
\begin{array}{l}
\displaystyle{%
   i\hbar\,\frac{\partial\psi}{\partial t}-U(x)\,\psi = \sqrt{m^2c^4-\hbar^2c^2\Delta_A}\,\,\psi,  %
}
\end{array}
\end{equation}
where $U(x)=-e\Phi(x)$ is the potential energy in an external scalar field. %

\section{Schr\"{o}dinger equation with the relativistic correction}\vspace{-0mm} %
Let us write the Schr\"{o}dinger equation with the scalar potential,
taking into account the main relativistic correction. In this approximation %
\begin{equation} \label{35}
\begin{array}{l}
\displaystyle{%
  \sqrt{1-\lambdabar^2\Delta}\,\,\psi\approx\bigg(1-\frac{1}{2}\,\lambdabar^2\Delta-\frac{1}{8}\,\lambdabar^4\Delta^2\bigg)\psi. %
}%
\end{array}
\end{equation}
Then equation (34) will take the form
\begin{equation} \label{36}
\begin{array}{l}
\displaystyle{%
   i\hbar\,\frac{\partial\psi}{\partial t}=U(x) + mc^2\bigg(1-\frac{1}{2}\,\lambdabar^2\Delta-\frac{1}{8}\,\lambdabar^4\Delta^2\bigg)\psi.  %
}
\end{array}
\end{equation}
In order to proceed to the nonrelativistic limit, it is convenient
to pass to the wave function $\chi(x)$, which differs from the
original function by a time-dependent phase factor
\begin{equation} \label{37}
\begin{array}{l}
\displaystyle{%
   \psi(x)=\chi(x)\,e^{-i\frac{mc^2}{\hbar}\,t}. %
}%
\end{array}
\end{equation}
Then equation (36) will take the form
\begin{equation} \label{38}
\begin{array}{l}
\displaystyle{%
   i\hbar\,\frac{\partial\chi}{\partial t}=-\frac{\hbar^2}{2m}\Delta\chi + U(x)\,\chi - \frac{\hbar^4}{8m^3c^2}\Delta^2\chi. %
}
\end{array}
\end{equation}
Here the last term is the relativistic correction. The probability
flux density (18) ${\bf j}={\bf j}^{(1)}+{\bf j}^{(2)}$ is the sum
of the nonrelativistic part ${\bf j}^{(1)}$ and the relativistic
correction ${\bf j}^{(2)}$:
\begin{equation} \label{39}
\begin{array}{ccc}
\displaystyle{\hspace{0mm}%
  {\bf j}^{(1)}=-\frac{i\hbar}{2m}\big(\chi^*\nabla\chi -\chi\nabla\chi^*\big),  %
}\vspace{2mm}\\ %
\displaystyle{\hspace{0mm}%
  {\bf j}^{(2)}=-\frac{i\hbar}{8m}\lambdabar^2\big(\chi^*\nabla\Delta\chi -\chi\nabla\Delta\chi^* +\Delta\chi^*\nabla\chi-  \Delta\chi\nabla\chi^* \big).  %
}%
\end{array}
\end{equation}

The energy density (22) and the energy flux density %
${\bf j}_E={\bf j}_E^{(1)}+{\bf j}_E^{(2)}$ (25) in the considered
approximation are given by the formulas
\begin{equation} \label{40}
\begin{array}{ccc}
\displaystyle{\hspace{0mm}%
  {\rm H}=mc^2|\chi|^2-\frac{\hbar^2}{4m}\big(\chi^*\Delta\chi +\chi\Delta\chi^*\big)-  %
  \frac{\hbar^2}{16m}\lambdabar^2\big(\chi^*\Delta^2\chi +\chi\Delta^2\chi^*\big),  %
}%
\end{array}
\end{equation}
\vspace{-4mm} %
\begin{equation} \label{41}
\begin{array}{ccc}
\displaystyle{\hspace{0mm}%
  {\bf j}_E^{(1)}=\frac{\hbar^2}{4m}\big(\dot{\chi}^*\nabla\chi + \dot{\chi}\nabla\chi^*-\chi^*\nabla\dot{\chi}-\chi\nabla\dot{\chi}^* + %
  i\frac{\hbar}{2m}mc^2\big(\chi^*\nabla\chi-\chi\nabla\chi^*\big), %
}\vspace{4mm}\\ %
\displaystyle{\hspace{0mm}%
  {\bf j}_E^{(2)}=\frac{\hbar^2}{16m}\lambdabar^2\Big[ %
  \big(\nabla\chi^*\big)\big(\Delta\dot{\chi}\big)+\big(\Delta\dot{\chi}^*\big)\big(\nabla\chi\big)+
  \big(\nabla\Delta\chi^*\big)\dot{\chi}+\dot{\chi}^*\big(\nabla\Delta\chi\big)-  %
}\vspace{2mm}\\ %
\displaystyle{\hspace{26mm}%
  -\chi^*\big(\nabla\Delta\dot{\chi}\big)-\big(\nabla\Delta\dot{\chi}^*\big)\chi
  -\big(\Delta\chi^*\big)\big(\nabla\dot{\chi}\big)-\big(\nabla\dot{\chi}^*\big)\big(\Delta\chi\big)\Big]-   %
}\vspace{2mm}\\ %
\displaystyle{\hspace{13mm}%
  -i\frac{\hbar^2}{8m}mc^2\lambdabar^2\Big[ %
  \big(\nabla\chi^*\big)\big(\Delta\chi\big)-\big(\Delta\chi^*\big)\big(\nabla\chi\big)+   %
  \big(\nabla\Delta\chi^*\big)\chi - \chi^*\big(\nabla\Delta\chi\big) \Big]. %
}%
\end{array}
\end{equation}
As we see, in the continuity equation for energy, even in the
nonrelativistic limit, it is necessary to take into account the rest
energy of the particle.

The momentum density (27) is expressed through the probability flux
density in the nonrelativistic approximation by the relation $\boldsymbol{\pi}=m{\bf j}^{(1)}$. %
The flux density of  $i$\,-\,component of momentum $\sigma_{ik}$ is given by the formulas %
\begin{equation} \label{42}
\begin{array}{ccc}
\displaystyle{\hspace{0mm}%
  \sigma_{ik}=\sigma_{ik}^{(1)} + \sigma_{ik}^{(2)},  %
}\vspace{3mm}\\ %
\displaystyle{\hspace{0mm}%
  \sigma_{ik}^{(1)}=\frac{\hbar^2}{4m}\Big[ %
  \chi^*\big(\nabla_i\nabla_k\,\chi\big) + \chi\big(\nabla_i\nabla_k\,\chi^*\big)-\nabla_i\chi^*\nabla_k\chi - \nabla_i\chi\nabla_k\chi^*\Big],  %
}\vspace{3mm}\\ %
\displaystyle{\hspace{26mm}%
  \sigma_{ik}^{(2)}=\frac{\hbar^2}{16m}\lambdabar^2\Big[ %
  \chi^*\big(\nabla_i\nabla_k\Delta\chi\big) + \chi\big(\nabla_i\nabla_k\Delta\chi^*\big)+ %
  \Delta\chi^*\big(\nabla_i\nabla_k\Delta\chi\big)+\Delta\chi\big(\nabla_i\nabla_k\Delta\chi^*\big)-  %
}\vspace{2mm}\\ %
\displaystyle{\hspace{45mm}%
  -\big(\nabla_i\Delta\chi^*\big)\nabla_k\chi-\big(\nabla_k\Delta\chi^*\big)\nabla_i\chi  %
  -\nabla_i\chi^*\big(\nabla_k\Delta\chi\big)-\nabla_k\chi^*\big(\nabla_i\Delta\chi\big)\Big]. %
}%
\end{array}
\end{equation}
In equation (38) the last term should be considered as a small perturbation. %

\section{Evolution of the wave function in an unlimited space}\vspace{-0mm} %
Let us consider the evolution of the wave function in an unlimited
medium. Let us move on to the new wave function $\chi(x)$ (37), such
that $\psi(x)=\chi(x)\exp\!\big(\!-i\,t/\tau_C\big)$, where
$\tau_C\equiv\lambdabar/c=\hbar/mc^2$. Note that for an electron
$\tau_C\approx 1.3\cdot10^{-21}$\,s. In this case equation (6) will take the form %
\begin{equation} \label{43}
\begin{array}{l}
\displaystyle{%
   i\tau_C\dot{\chi}=\Big(\!\sqrt{1-\lambdabar^2\Delta}-1\Big)\chi\,. %
}
\end{array}
\end{equation}
The equations for the real and imaginary parts of the wave function %
$\chi=\big(\chi'+i\chi''\big)\big/\sqrt{2}$ can be written in the form %
\begin{equation} \label{44}
\begin{array}{l}
\displaystyle{%
   \tau_C\dot{\chi'}=\Big(\!\sqrt{1-\lambdabar^2\Delta}-1\Big)\chi'', \quad  %
   \tau_C\dot{\chi''}=-\Big(\!\sqrt{1-\lambdabar^2\Delta}-1\Big)\chi'.  %
}%
\end{array}
\end{equation}
There holds the normalization condition
\begin{equation} \label{45}
\begin{array}{l}
\displaystyle{%
   \frac{1}{2}\int\!d{\bf x}\Big[\chi'^2({\bf x},t)+\chi''^2({\bf x},t)\Big]=1.  %
}%
\end{array}
\end{equation}

Let us consider the evolution of the wave function in an unlimited
space, provided that at the initial moment $t=0$ there are given the
functions $\chi'({\bf x},0)$ and $\chi''({\bf x},0)$, normalized by
the condition (45) . From equations (44), it follows a second order
equation in time for the function $\chi'$:
\begin{equation} \label{46}
\begin{array}{l}
\displaystyle{%
   \tau_C^2\ddot{\chi'}+\Big(\!\sqrt{1-\lambdabar^2\Delta}-1\Big)^2\chi'=0.  %
}%
\end{array}
\end{equation}
We expand the required functions into the Fourier integral
\begin{equation} \label{47}
\begin{array}{l}
\displaystyle{%
   \chi'({\bf x},t)=\frac{1}{(2\pi)^{3/2}}\int\!\chi'_k(t)\,e^{i{\bf k}{\bf x}}d{\bf k},  \quad  %
   \chi''({\bf x},t)=\frac{1}{(2\pi)^{3/2}}\int\!\chi''_k(t)\,e^{i{\bf k}{\bf x}}d{\bf k}.  %
}%
\end{array}
\end{equation}
Substituting these expansions into (46) and the second equation (44), we obtain %
\begin{equation} \label{48}
\begin{array}{l}
\displaystyle{%
   \ddot{\chi'}+\omega_k^2\chi'_k=0, \qquad  \dot{\chi''}+\omega_k\chi'_k=0,  %
}%
\end{array}
\end{equation}
where
\begin{equation} \label{49}
\begin{array}{l}
\displaystyle{%
   \omega_k=\Big(\!\sqrt{1+\lambdabar^2k^2}-1\Big)\frac{1}{\tau_C}.  %
}%
\end{array}
\end{equation}
In the nonrelativistic limit $\lambdabar k\ll 1$
\begin{equation} \label{50}
\begin{array}{l}
\displaystyle{%
   \omega_k\approx\frac{\lambdabar^2k^2}{2\tau_C}=\frac{\hbar k^2}{2m}.  %
}%
\end{array}
\end{equation}
The group velocity, which determines the speed of propagation of a
wave-particle
\begin{equation} \label{51}
\begin{array}{l}
\displaystyle{%
   \nu_g=\frac{d\omega_k}{dk}=\lambdabar\,\frac{ck}{\sqrt{1+\lambdabar^2k^2}},  %
}%
\end{array}
\end{equation}
cannot exceed the speed of light. A calculation of the group
velocity within the framework of ordinary nonrelativistic quantum
mechanics by formula (50) gives $\nu_g=\hbar k/m$, so that with an
increase of $k$ or a decrease of the de Broglie wavelength the speed
of signal propagation can be arbitrarily large.

Solutions of equations (48) have the form
\begin{equation} \label{52}
\begin{array}{l}
\displaystyle{%
   \chi'_k(t)=A_ke^{-i\omega_kt} + B_ke^{i\omega_kt}, \qquad  %
   \chi''_k(t)=-iA_ke^{-i\omega_kt} + iB_ke^{i\omega_kt}.  %
}%
\end{array}
\end{equation}
From the conditions for reality of the functions $\chi',\chi''$ it
follows $\chi'_k=\chi'^{\,*}_{-k}$ and $\chi''_k=\chi''^{\,*}_{-k}$, %
so that $B_k=A_{-k}^{\,*}$. The normalization condition (45) gives
\begin{equation} \label{53}
\begin{array}{l}
\displaystyle{%
   \int\!d{\bf k}\,\Big(|A_k|^2+|A_{-k}|^2\Big)=1.  %
}%
\end{array}
\end{equation}
The coefficients in (52) can be expressed through the initial conditions %
\begin{equation} \label{54}
\begin{array}{l}
\displaystyle{%
   A_k=\frac{1}{2}\big(\chi'_k(0)+i\chi''_k(0)\big),   \quad  %
   A_{-k}^{\,*}=\frac{1}{2}\big(\chi'_k(0)-i\chi''_k(0)\big),  %
}%
\end{array}
\end{equation}
so that
\begin{equation} \label{55}
\begin{array}{ccc}
\displaystyle{\hspace{0mm}%
  \chi'_k(t)=\chi'_k(0)\cos\omega_kt + \chi''_k(0)\sin\omega_kt,  %
}\vspace{4mm}\\ %
\displaystyle{\hspace{0mm}%
  \chi''_k(t)=-\chi'_k(0)\sin\omega_kt + \chi''_k(0)\cos\omega_kt\,.  %
}%
\end{array}
\end{equation}
Taking into account the relations
\begin{equation} \label{56}
\begin{array}{l}
\displaystyle{%
   \chi'_k(0)=\frac{1}{(2\pi)^{3/2}}\int\!d{\bf x}\,e^{-i{\bf k}{\bf x}}\chi'({\bf x},0),  \quad  %
   \chi''_k(0)=\frac{1}{(2\pi)^{3/2}}\int\!d{\bf x}\,e^{-i{\bf k}{\bf x}}\chi''({\bf x},0),       %
}%
\end{array}
\end{equation}
we find solutions which determine the real and imaginary parts of
the wave function at an arbitrary moment of time from their values
at the initial moment of time
\begin{equation} \label{57}
\begin{array}{ccc}
\displaystyle{\hspace{0mm}%
  \chi'({\bf x},t)=\int\!d{\bf x}'\big[J_C({\bf x}-{\bf x}',t)\chi'({\bf x}',0)+J_S({\bf x}-{\bf x}',t)\chi''({\bf x}',0)\big],  %
}\vspace{2mm}\\ %
\displaystyle{\hspace{0mm}%
  \chi''({\bf x},t)=\int\!d{\bf x}'\big[-J_S({\bf x}-{\bf x}',t)\chi'({\bf x}',0)+J_C({\bf x}-{\bf x}',t)\chi''({\bf x}',0)\big],  %
}%
\end{array}
\end{equation}
where
\begin{equation} \label{58}
\begin{array}{l}
\displaystyle{%
   J_C({\bf x}-{\bf x}',t)=\frac{1}{(2\pi)^3}\int\!d{\bf k}\cos\omega_kt\,e^{i{\bf k}({\bf x}-{\bf x}')}, \quad  %
   J_S({\bf x}-{\bf x}',t)=\frac{1}{(2\pi)^3}\int\!d{\bf k}\sin\omega_kt\,e^{i{\bf k}({\bf x}-{\bf x}')}.        %
}%
\end{array}
\end{equation}
Using the expansion of a plane wave in terms of spherical functions
and integrating over angles, we obtain
\begin{equation} \label{59}
\begin{array}{ccc}
\displaystyle{\hspace{0mm}%
   J_C({\bf x}-{\bf x}',t)\equiv J_C(|{\bf x}-{\bf x}'|,t)=\frac{1}{2\pi^2}\int_0^\infty\!dk\,k^2\cos(\omega_kt)\,j_0\big(k|{\bf x}-{\bf x}'|\big),   %
}\vspace{2mm}\\ %
\displaystyle{\hspace{0mm}%
   J_S({\bf x}-{\bf x}',t)\equiv J_S(|{\bf x}-{\bf x}'|,t)=\frac{1}{2\pi^2}\int_0^\infty\!dk\,k^2\sin(\omega_kt)\,j_0\big(k|{\bf x}-{\bf x}'|\big),   %
}%
\end{array}
\end{equation}
where $j_0(x)=\sin x/x$ is the spherical Bessel function.

\begin{figure}[b!]
\vspace{0mm}  \hspace{0mm}
\includegraphics[width = 7.8cm]{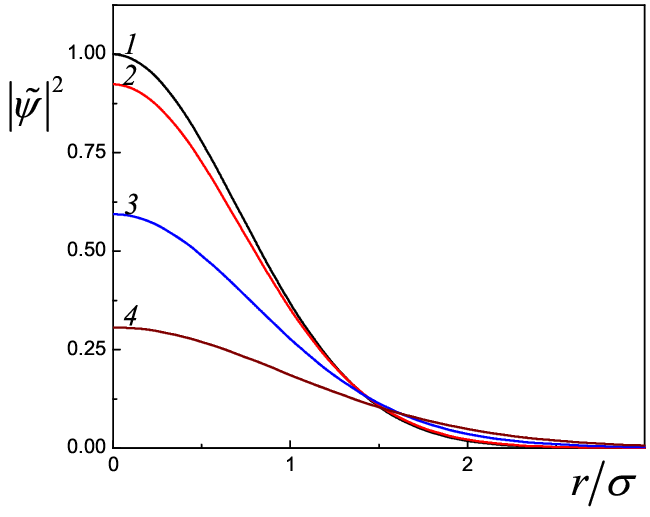} 
\vspace{-3mm} %
\caption{\label{fig01} 
The spatial distribution of the probability density
$|\tilde{\psi}|^2\equiv\pi^{3\!/2}\sigma^3|\psi|^2$ for $\sigma/\lambdabar=1$ %
at the moments of time $\tau=t/\tau_C:$ %
{\it 1} -- 0;\, {\it 2} -- 0.5;\, {\it 3} -- 1.3;\, {\it 4} -- 2.0.  %
}%
\end{figure}

Let us express the sought functions at the initial moment $t=0$ in
terms of the modulus and phase
\begin{equation} \label{60}
\begin{array}{l}
\displaystyle{%
   \chi'({\bf x},0)=\sqrt{2}\,|\psi({\bf x},0)|\cos\theta({\bf x},0), \quad %
   \chi''({\bf x},0)=\sqrt{2}\,|\psi({\bf x},0)|\sin\theta({\bf x},0).        %
}%
\end{array}
\end{equation}
As the initial condition we choose the Gaussian distribution for the
probability density
\begin{equation} \label{61}
\begin{array}{l}
\displaystyle{%
   |\psi(r,0)|^2=\frac{1}{\pi^{3/2}\sigma^3}\,e^{-r^2\!/\sigma^2},       %
}%
\end{array}
\end{equation}
where $r\equiv|{\bf x}|$, and the parameter $\sigma$ determines the
width of the wave packet, so that as $\sigma$ decreases the
distribution (61), satisfying the normalization condition %
$4\pi\!\int_0^\infty|\psi(r,0)|^2r^2dr=1$, approaches the delta
function. For the initial phase in (60) we choose $\theta({\bf x},0)=0$, %
so that for the real and imaginary parts we have
\begin{equation} \label{62}
\begin{array}{l}
\displaystyle{%
   \chi'(r,0)=\sqrt{2}\,|\psi(r,0)|=\frac{\sqrt{2}}{\pi^{3/4}\sigma^{3/2}}\,e^{-r^2\!/(2\sigma^2)}, \quad  \chi''(r,0)=0.    %
}%
\end{array}
\end{equation}
In the calculation it is helpful to use the well-known integral [6]
\begin{equation} \label{63}
\begin{array}{l}
\displaystyle{%
   \int_0^\infty e^{-a^2t^2}t^{\nu+3\!/2}j_{\nu-1\!/2}(bt)\,dt = \sqrt{\frac{\pi}{2}}\frac{b^{\nu-1\!/2}}{(2a^2)^{\nu+1}}\,e^{-\frac{b^2}{4a^2}}.    %
}%
\end{array}
\end{equation}
As a result, we obtain a solution in the form
\begin{equation} \label{64}
\begin{array}{ccc}
\displaystyle{\hspace{0mm}%
   \chi'(r,t)=\frac{\sqrt{2}}{\pi^{3\!/4}}\sqrt{\frac{2}{\pi}}\,\sigma^{3\!/2}\int_0^\infty dk\,k^2\cos(\omega_kt)\,j_0(kr)\,e^{-\frac{k^2\sigma^2}{2}},     %
}\vspace{2mm}\\ %
\displaystyle{\hspace{0mm}%
   \chi''(r,t)=-\frac{\sqrt{2}}{\pi^{3\!/4}}\sqrt{\frac{2}{\pi}}\,\sigma^{3\!/2}\int_0^\infty dk\,k^2\sin(\omega_kt)\,j_0(kr)\,e^{-\frac{k^2\sigma^2}{2}}.    %
}%
\end{array}
\end{equation}
The spatial distribution of the probability density
$|\tilde{\psi}|^2\equiv\pi^{3\!/2}\sigma^3|\psi|^2=\pi^{3\!/2}\sigma^3|\chi|^2$ %
at some moments of time is shown in Figure 1. %
As we see, a particle localized at the initial moment of time at the
origin of coordinate system is gradually with equal probability
spreading throughout space.

\section{Relativistic correction in the oscillator theory}\vspace{-0mm} %
The Schr\"{o}dinger equation for a one-dimensional along the
$x$\,-\,axis quantum oscillator with account of the main
relativistic correction (38) has the form
\begin{equation} \label{65}
\begin{array}{l}
\displaystyle{%
   i\hbar\frac{\partial\chi}{\partial t}=-\frac{\hbar^2}{2m}\frac{\partial^2\chi}{\partial x^2}+ %
   \frac{m\omega^2x^2}{2}\,\chi-\frac{\hbar^4}{8m^3c^2}\frac{\partial^4\chi}{\partial x^4}.    %
}%
\end{array}
\end{equation}
In the stationary case, when $\chi(x,t)=\varphi(x)\exp\!\big(\!-iEt/\hbar\big)$, we have %
\begin{equation} \label{66}
\begin{array}{l}
\displaystyle{%
   E\varphi = -\frac{\hbar^2}{2m}\frac{d^2\varphi}{d x^2}+ %
   \frac{m\omega^2x^2}{2}\,\varphi-\frac{\hbar^4}{8m^3c^2}\frac{d^4\varphi}{d x^4}.    %
}%
\end{array}
\end{equation}
It is convenient to introduce a dimensionless coordinate $\xi\equiv\big(\!\sqrt{m\omega/\hbar}\big)x$, %
and in result the equation takes the form
\begin{equation} \label{67}
\begin{array}{l}
\displaystyle{%
   \varepsilon\varphi = -\varphi^{(I\!I)}+\xi^2\varphi-\gamma\varphi^{(I\!V)}, %
}%
\end{array}
\end{equation}
where differentiation occurs with respect to $\xi$, %
$\varepsilon\equiv2E/\hbar\omega$, and
\begin{equation} \label{68}
\begin{array}{l}
\displaystyle{%
  \gamma\equiv\frac{\hbar\omega}{4mc^2}.   %
}%
\end{array}
\end{equation}
In order to estimate this dimensionless parameter, we take %
$\hbar\omega\sim1\,{\rm K}\approx10^{-16}$\,erg, $m\approx10^{-23}$\,g.  %
As a result, we get a very small value $\gamma\sim10^{-14}$.

The correction to $n$\,-\,th energy level is determined by the
diagonal element of the perturbation operator taken over the
unperturbed wave functions [5]
\begin{equation} \label{69}
\begin{array}{l}
\displaystyle{%
  \varepsilon^{(1)}=-\gamma\big\langle n\big|\frac{d^4}{d\xi^4}\big|n\big\rangle.    %
}%
\end{array}
\end{equation}
The wave functions of an oscillator in the coordinate representation
are expressed through Hermite polynomials [5]
\begin{equation} \label{70}
\begin{array}{l}
\displaystyle{%
  \big\langle \xi\big|n\big\rangle = \bigg(\frac{m\omega}{\pi\hbar}\bigg)^{\!1\!/4}\frac{1}{\sqrt{2^nn!}}\,e^{-\frac{\xi^2}{2}}H_n(\xi).    %
}%
\end{array}
\end{equation}
As a result, we find
\begin{equation} \label{71}
\begin{array}{l}
\displaystyle{%
  \varepsilon^{(1)}=-\,\gamma\,\frac{3}{2}\,\big(n+1\big)\big(n+2\big).   %
}%
\end{array}
\end{equation}
Thus, the energy of the oscillator level with account of the main
relativistic correction takes the form
\begin{equation} \label{72}
\begin{array}{l}
\displaystyle{%
  E_n=\hbar\omega\bigg[n\bigg(1-\frac{9}{4}\,\gamma\bigg)+\frac{1}{2}\big(1-3\gamma\big)-\frac{3}{4}\,\gamma n^2\bigg].   %
}%
\end{array}
\end{equation}
As we see, the relativistic effect leads to a decrease of the ground
and excited levels, as well as to a violation of the equidistance of
the spectrum. The effect of violation of the equidistance property,
leading to a decrease in the distance between higher levels, can be
used for the experimental detection of relativistic effects in
quantum mechanics.

\section{Conclusion}\vspace{-2mm} %
The paper proposes the relativistic generalization of the
Schr\"{o}dinger equation for the complex function, which can be
interpreted as a probability amplitude. Unlike the nonrelativistic
equation, this equation allows propagation of signals only at a
speed not exceeding the speed of light. The theory is also
formulated within the Lagrangian formalism. The equations of
conservation of the probability, energy and momentum of the complex
field are obtained. The problem of evolution of a wave packet in an
unlimited space is solved. The relativistic corrections to the
energy levels of the harmonic oscillator are found. It is shown that
relativistic effects lead to a shift and a violation of the level
equidistance, which can be used for experimental detection of the
relativistic effect in the probabilistic quantum theory. %

The author thanks A.A. Soroka for help in preparing the article.


\vspace{5mm} %

\end{document}